\documentstyle[twocolumn,aps,epsfig,floats]{revtex}

\begin{document}

\draft \tolerance = 10000

\setcounter{topnumber}{1}
\renewcommand{\topfraction}{0.9}
\renewcommand{\textfraction}{0.1}
\renewcommand{\floatpagefraction}{0.9}

\twocolumn[\hsize\textwidth\columnwidth\hsize\csname
@twocolumnfalse\endcsname

\title{Is it Possible to Describe Economical Phenomena \\ by Methods of Statistical Physics of Open Systems?}
\author{L.Ya. Kobelev, O.L. Kobeleva, Ya.L. Kobelev \\
Department of Problems of Extreme Influence on Matter, \\ Department  of
Low Temperatures Physics, Ural State University,\\ Lenina av., 51,
Ekaterinburg, 620083, RUSSIA}

\maketitle
\begin{abstract}
The methods of statistical physics of open systems are used for describing
the time dependence of economic characteristics (income, profit, cost,
supply, currency etc.) and their correlations with each other. Nonlinear
equations ( analogies of known reaction-diffusion, kinetic, Langevin
equation) describing appearance of bifurcations, self-sustained
oscillational processes, self-organizations in economic phenomena are
offered.

\end{abstract}
\vspace{1cm}

]

\section{Introduction}
One of  the  systems consisting of a big number of elements, which
exchange information, currency, goods etc. with each other is a market
economy. The number of the elements ( for example, of firms characterized
by income, dividend and investments size, demand and supply value, labor
cost and so on) is so great that application of statistical physics
methods to describe market economy becomes possible. But now (as it seems)
there is no  a general algorithms for  subsequent theories applying
methods of statistical physics of open systems to market economy (although
G.Bystrai \cite{Byst1}-\cite{Byst3},A.Johansen and D.Sernette
\cite{Sorn},D.Meodows \cite{Meod}, W.Weidlich \cite{Weid}, I.Lubashevsky
\cite{Luba} used some ideas of fractal geometry and nonequilibrium
thermodynamics). In the present paper the methods of constructing of
equations to describe economy characters behaviour are stated (see
\cite{Kob1}) in the frames of mathematical formalism developed in
statistical physics and describing both self-organization processes and
catastrophes. To do this a number of economic concepts $x_i(t)$
($i=1,...N$; $t$ denotes time) is introduced and used, different for
different economic problems, with $x_i$ being able to depend on other
economic characteristics as well. The concepts mentioned are then treated
as "coordinates" of an economic system. Another set of economic variables
$F_\alpha(x_1...x_N,t,\dot{x}_1...\dot{x}_N)$ is defined as functions of
$x_i(t)$, $\dot{x}_i(t)$, $\frac{\partial x_i}{\partial \chi_k}$
($i,j=1...N, \alpha=1...\beta$) etc. The set of "variables" $x_i$ and
their relations with $F_\alpha$ are relative, that is the "coordinates"
$x_i$ in a concrete problem can turn to be functions $F_\alpha$ in another
one. If we are to regard $F_\alpha(x_i(t), \dot{x}_i(t),\frac{\partial
x_i}{\partial\chi_j},t)$ for a given $\alpha$ (i.e., $F_\alpha$ is a
currency rate of exchange) as a set of macroscopic functions
characterizing the economic system under consideration, it is possible to
write down well developed in statistical physics of open systems equations
of nonlinear kinetic, reaction-diffusion, oscillational types. Treatment
of these equations and their solutions, if being applied to a concrete
economic problem, allows to describe not only well known economic
regularities, but also to define the values of  and relations between
economic parameters, leading to drastic violations of an equilibrium.
\section{Formulation of the equations.}
1. Consider an economic media consisting of $N$ elements $x_i$
($i=1,...N$). Let this elements define functions  $F_\alpha$ (not
necessary independent ones). If while changing governing parameters the
elements themselves change (an active media), then, provided the diffusion
type relations presence, the time behaviour of $F_\alpha(x_i(t),
\dot{x}_i(t),\frac{\partial x_i}{\partial \chi_j},t$ set can be described
by a reaction diffusion equations
\begin{eqnarray}\label{1}  \nonumber&& \frac{dF_\alpha(x_i,t)}{dt}=\varphi_\alpha(F)+\frac{\partial}{\partial
x_i}[D_{ij}^\alpha(F)\frac{\partial F_\alpha}{\partial x_j}-A_i(x,t)
F_\alpha]+ \\ &&+\frac{\partial}{\partial\dot{x}_i}[\tilde
D_{ij}^\alpha(F)\frac{\partial F_\alpha}{\partial \dot{x}_j}+
(a_i-b_i(x,t)F_\alpha]
\end{eqnarray}
$\alpha=1...\beta$, $i,j=1...N$, $\beta$ is the number of $F_a$ functions
used in the problem under investigation. The full time derivative in the
left-hand side of \ref{1} can be represented as (summing up over the
repeating indexes)
\begin{equation}\label{2}
\frac{dF_\alpha}{dt}=\frac{\partial F_\alpha}{\partial t}+\frac{\partial
F_\alpha}{\partial x_\beta}\cdot\frac{dx_\beta}{dt}+\frac{\partial
F_\alpha}{\partial\dot{x}_\beta}\cdot\frac{d\dot{x}_\beta}{dt}
\end{equation}
Thus, the system of equations describing an economic situation reads
\begin{equation}\label{3}
\frac{\partial F_\alpha}{\partial t}+\frac{\partial F_\alpha} {\partial
x_\beta}\cdot\frac{dx_\beta}{dt}+\frac{\partial F_\alpha}
{\partial\dot{x}_\beta}\cdot\frac{d\dot{x}_\beta}{dt}=I_\alpha
\end{equation}
where
\begin{eqnarray}\label{4} \nonumber
I_\alpha=&&\varphi(F_i,\dot{F}_i...t)+\frac{\partial}{\partial
x_i}[D_{ij}^\alpha(F)\frac{\partial F_\alpha}{\partial x_j}-A_i(x,t)
F_\alpha]+ \\   &&+\frac{\partial}{\partial\dot{x}_i}[\tilde
D_{ij}^\alpha(F)\frac{\partial F_\alpha}{\partial \dot{x}_j}+
(a_i-b_i(x,t)F_\alpha]
\end{eqnarray}
Here $D_{ij}^\alpha(F_1...F_\beta)$, $\tilde{D}_{ij}^\alpha
(F_1...F_\beta)$ are the diffusion coefficients in the $x_i$ and
$\dot{x}i$ - spaces, depending (including nonlinear dependencies) on $F$;
$\varphi(F_i,\dot{F}_i...t)$ - is a nonlinear function, determined by the
economic problem data (as well as the other parameters dependencies). In
(\ref{3}) $\dot{x}_i$ and $\ddot{x}_i$ characterize the rate of and the
acceleration of economic element $x_b$ changing with time; $\dot{x}_i$ and
$\frac{d\dot{x}_i}{dt}$ can be determined here by their own equations
\begin{eqnarray}\label{4ab} \nonumber
&& a)\;\frac{d\dot{x}}{dt}=\tilde{\varphi}_1(F_\alpha,B(x_i,x_j)) \\
\nonumber &&
b)\;\frac{d^2\dot{x}}{dt^2}=\tilde{\varphi}_2(F_\alpha,A(x_i,x_j))
\end{eqnarray}
where $A$ and $B$ are "external" influences onto the economic system). In
special cases $I_\alpha$ may equals zero $I_\alpha=0$ for  some or all of
$\alpha$ (it corresponds neglecting the influence of the factors promoting
$F_\alpha$ varying with time).

2. Above there were formulated equations (\ref{1})-(\ref{3}) describing
time dependence of economic parameters (considered as variables or
functions) of an open non equilibrium statistical system and including the
description of crisis or self-sustain oscillation processes as well.
Nevertheless, so far as variables $x_i$ themselves can depend on the other
variables $x_j$ ($i\neq j$) and on functions $F_\alpha(x_i...)$ (stress
again that in economy, because of the dependence of almost all the factors
on each other, the choice of variables $x_i$ and functions
$F_\alpha(x_i...)$ is quite an arbitrary and may change from one problem
to another), one should supplement  the eqs. (\ref{1})-(\ref{3}) by a
system of equations describing the $x_i$ dependence on $x_j$ and
$F_\alpha$. These equations can be chosen basing on the same ideas at the
time equations, that is can be obtained by substitution of
$\frac{\partial}{\partial t}$ for $\frac{\partial} {\partial x_i}$ or for
$\frac{\partial}{\partial F_\alpha}$. In the case of reaction-diffusion
equations one then writes:
\begin{eqnarray}\label{5}  \nonumber
\frac{dx_i}{dx_j} &=& F_{ij}(x_j,F_\alpha)+\frac{\partial}{\partial x_j}
(D_{j\beta}^x(x_j)\frac{\partial}{\partial x_\beta}\cdot x_i-A_j x_i)+ \\
 &+& \frac{\partial}{\partial\dot{F}_\alpha}[\tilde D_{\gamma
\beta}(\dot{x})\frac{\partial}{\partial\dot{F}_\beta}\cdot x_i+(\tilde a_i
-\tilde b_i(x,t)x_\alpha)]
\end{eqnarray}
\begin{eqnarray}\label{6}  \nonumber
\frac{dx_i}{dF_\alpha} &=& \tilde{F}_{i\alpha}+\frac{\partial}{\partial
F_\gamma}[(D_{\gamma\beta}(x_j)\frac{\partial}{\partial F_\beta} \cdot
x_i-A_i x_\alpha)]+ \\  &+& \frac{\partial}{\partial \dot{x}_j}
[\tilde{D}_{j\beta }^{\dot{x}}(\dot{x}) \frac{\partial}{\partial
\dot{x}_\beta} \cdot x_i+(\tilde{a}_j-\tilde{b}_j(x,t)x_i)]
\end{eqnarray}

3. One can include in eqs. (\ref{5})-(\ref{6}) terms corresponding to
"diffusion" in the "velocity" space. The sums of such "diffusion" terms
should be added to the right-hand part of these equations. The system of
equation (\ref{1})-(\ref{6}) seems to be able to give a full enough
description of an economic system behaviour near bifurcation points and
phase transitions.

4. Let $I_a$ has the following form (here we neglect by diffusion and
nonlinear with respect to $F$ terms),
\begin{eqnarray}\label{7}
\nonumber &&a)\; I_\alpha=0 \\ \nonumber
&&b)\;I_\alpha=\frac{f_\alpha(x)-F_{\alpha_0}}{\tau}
\end{eqnarray}
$\tau$ is the relaxation time.

Then equation (\ref{3}) reads
\begin{eqnarray}\label{8}  \nonumber
a)\; \frac{dF_\alpha}{dt}&=&\frac{\partial F_\alpha}{\partial t}+\frac
{\partial F_\alpha}{\partial x_\beta}\cdot\frac{dx_\beta}{dt}+\frac
{\partial F_\alpha}{\partial\dot{x}_\beta}\cdot\frac{d\dot{x}_\beta}{dt}=0
\\ b)\; \frac{dF_\alpha}{dt}&=&\frac{\partial F_\alpha}{\partial t}+\frac
{\partial F_\alpha}{\partial x_\beta}\cdot\frac{dx_\beta}{dt}+\frac
{\partial F_\alpha}{\partial\dot{x}_\beta}\cdot\frac{d\dot{x}_\beta}{dt}=
\\ \nonumber &=& \frac{F_\alpha(x)-F_{\alpha _0}}{\tau}
\end{eqnarray}
Eq. (\ref{8}a) corresponds to a stationary  behaviour of $F_\alpha$ and
the time dependence is entirely defined by time dependencies of $x_\beta$
and $\dot{x}_\beta$. The case of (\ref{8}b) characterize the temporal
behaviour of the system deviation from the equilibrium state
$F_{\alpha_0}(x^0,t)$. The relaxation time can be a function of $F_\alpha$
(a nonlinear case).

The case of $\varphi=\alpha_0 F+\alpha_1 F^3$ includes nonlinear equations
of Landau-Ginzburg type.
\section{Statistical description of economics phenomenon by kinetic equations}
Equations (\ref{3})-(\ref{4}) govern the  behaviour of functions $F_a$
which are set by a $x_i$ values. Introduce therefore a set of values taken
by every of the variables $x_i$. Let them be distributed in a random way.
Then introduce a function $f_\alpha(x_i,...,t)$ of density of probability
to find at the time of t values of  the economic parameters to be equal to
$x_i$ ($i=1...N$) (here we use one and the same symbol $x_i$ for an
economic parameter and its value). For the mean values we shall have
\begin{eqnarray}\label{9}  \nonumber
\langle F_\alpha \rangle=\int{F_\alpha(x_i...)f_\alpha(x_i)dx_i}
\end{eqnarray}
and the equations for $f_\alpha$ chose as
\begin{eqnarray}\label{10}
\frac{df_\alpha}{dt}=I_\alpha
\end{eqnarray}
where $I_\alpha$ and $\frac{df_\alpha}{dt}$ are as follows:
\begin{equation}\label{11}
\frac{df_\alpha}{dt}=\frac{\partial f_\alpha}{\partial t}+\frac{\partial
f_\alpha}{\partial i} \cdot \frac{dx_i}{dt}+\frac{\partial f_\alpha}
{\partial \dot{x}_i} \cdot \frac{d\dot{x}_i}{dt}
\end{equation}
\begin{eqnarray}\label{12}
I_\alpha&=&\varphi_\alpha(f,t)+\frac{\partial}{\partial x_i}[D_{ij}^\alpha
(f)\frac{\partial f_\alpha}{\partial x_j}-A_i(x,t)f_\alpha]+ \\ \nonumber
&+&\frac{\partial}{\partial \dot{x}_i}[\tilde D_{ij}^\alpha (f)\frac
{\partial f_\alpha}{\partial \dot{x}_j}+(a_i-b_i(x,t)f_\alpha]
\end{eqnarray}
These equations corresponds to the statistical description of economic
systems in terms of kinetic equations (well known in the theory of open
systems \cite{Klim}) for the economic parameters distribution function and
allow to calculate the mean values for economic functions  $F_a$.
      Adduce now one more method of statistical description of economic
parameters $x_i$ and $F_a$ based on using of equation like that of
Brownian motion in thermostat. The role of the "thermostat" is played in
economy by external influences (e.g. currency infusion and so on)
supporting any of the system parameters to be constant (currency rate of
exchange in our case correspondingly). Then the equation for $F_a$ is of
Langevin type and for the only function $F$ has the form
\begin{equation}\label{13}
\frac{d^2 F}{dt^2}+\gamma(F)\frac{dF}{dt}+\beta(F)=Y(t)
\end{equation}
where $\gamma(t)$ and $\beta(t)$ are nonlinear with respect to $F$
coefficients, $Y(t)$ is a random process corresponding, e.g., to a random
influence of prices, demand etc.\\ The processes described by (\ref{13})
can also be described in terms of the distribution function $f(F,x,t)$ and
equations  like (\ref{10})-(\ref{12}).
\section{Economical phenomenon in the multifractal economical spaces}
Many facts show on the possibility of useful describing of  economical
phenomenon \cite{Sorn}and phenomenon of social life
\cite{Kob2},\cite{Sorn} by using idea of fractal dimensions. Thus we point
out  the way of generalize of equations of above paragraphs for
multifractal economical spaces. For including multifractal characteristics
in equations (\ref{1})-(\ref{13}) we must use generalized
Riemann-Liouville derivatives defined in \cite{Kob3}
\begin{equation} \label{14}
D_{+,t}^{d}f(t)=\left( \frac{d}{dt}\right)^{n}\int_{a}^{t}
\frac{f(t^{\prime})dt^{\prime}}{\Gamma
(n-d(t^{\prime}))(t-t^{\prime})^{d(t^{\prime})-n+1}}
\end{equation}
\begin{equation} \label{15}
D_{-,t}^{d}f(t)=\left( \frac{d}{dt}\right)
^{n}\int_{t}^{b}\frac{(-1)^{n}f(t^{\prime})dt^{\prime}}{\Gamma
(n-d(t^{\prime}))(t^{\prime}-t)^{d(t^{\prime})-n+1}}
\end{equation}
where $\Gamma(x)$ is Euler's gamma function, and $a$ and $b$ are some
constants from $[0,\infty)$. In these definitions, as usually, $n=\{d\}+1$
, where $\{d\}$ is the integer part of $d$ if $d\geq 0$ (i.e. $n-1\le
d<n$) and $n=0$ for $d<0$. If $d=const$, the generalized fractional
derivatives (GFD) (\ref{14})-(\ref{15}) coincide with the Riemann -
Liouville fractional derivatives ($d\geq 0$) or fractional integrals
($d<0$). In the equations (\ref{1})-(\ref{13})for taking into account
multifractal characteristics of economics phenomenon it is necessary
change ordinary derivatives by generalized Riemann-Liouville derivatives.
\section{Conclusion}
 The equations given above are quite general and witness only a possibility
of describing economic processes in terms of the mathematical formalism
used in statistical physics of open systems in the economical (topological
or fractal) spaces . Taking into account that in market economy one can
observe catastrophes (e.g.,financial crises, bankrupts, fast changes in
the current exchange rate) in macro and microeconomic and stable processes
like self regulations, and the processes of development (stagnation,
demolishing) are determined by statistical reasons, using of the methods
of statistical physics given in the present paper may turn to be very
fruitful.  A general algorithm for economic processes description on the
basis of statistical physics equations for open system (mainly, equations
of reaction-diffusion type) is proposed. It contains wide opportunities to
use nonlinear equations and governing parameters for economic problems, so
far not developed in a systematic way. The methodics described can be
easily adjusted to study mathematical regularities in biology, medicine,
geology etc . We ask apology for absence  in this article of grate amount
of cites at economical  papers concerning the mathematical problems of
economics because we can not find direct connection  our description with
known description of economical problems. The authors thank Ph.D.
V.L.Kobeleva, the member of our research group, for useful and fruitful
discussion and grate help.

\end{document}